\documentclass[preprint,12pt]{elsarticle}
\usepackage{color,multirow}

\usepackage{graphics}

\usepackage{amssymb}

\usepackage{lineno}

\journal{NIM A }

\begin{document}

\begin{frontmatter}



\title{Sensitivity estimate of the MACE gamma ray telescope}


\author[lab1]{Mradul Sharma\footnote{\small mradul@barc.gov.in}}
\author[lab1]{Chinmay B}
\author[lab1]{Nilay Bhatt}
\author[lab1]{Subir Bhattacharyya}
\author[lab2]{S. Bose}
\author[lab1]{Abhas Mitra}
\author[lab1]{R. Koul}
\author[lab1]{A. K. Tickoo}
\author[lab1]{Ramesh C. Rannot}

\address[lab1]{Astrophysical Sciences Division, Bhabha Atomic Research Centre, Mumbai, India}
\address[lab2]{The Bayesian and Interdisciplinary Research Unit, Indian Statistical Institute,Kolkata, India}

\begin{abstract}
The MACE (Major Atmospheric Cherenkov Experiment) is a 21 m diameter $\gamma$-ray telescope which 
is presently being installed at Hanle in Ladakh, India (32$^0$  46$^{'}$ 46$^{"}$ N, 78$^0$ 58$^{'}$ 35$^{"}$ E) 
at an altitude of 4270 m a.s.l. Once operational, it will become the highest altitude very high energy (VHE) 
$\gamma$-ray telescope in the world based on Imaging Atmospheric Cherenkov Technique (IACT). In the present 
work, we discuss the sensitivity estimate of the MACE telescope by using a substantially large Monte Carlo 
simulation database at $5^0$ zenith angle. The sensitivity of MACE telescope is  estimated by carrying out 
the $\gamma$-hadron segregation using the Random Forest method. It is estimated that the MACE telescope will 
have an analysis energy threshold of 38 GeV for image intensities above 50 photoelectrons. 
The integral sensitivity for point like sources with Crab Nebula-like spectrum 
above 38 GeV is $\sim 2.7\%$ of Crab Nebula flux at 5 $\sigma$ statistical 
significance level in 50 hrs of observation.
\end{abstract}


\end{frontmatter}


\section{Introduction}
Among the various fields of astronomy (Radio, Infrared, Optical, Ultraviolet, X-rays, $\gamma$-rays, gravitational waves), 
ground based $\gamma$-ray astronomy is one of the youngest entrants. 
This field was pioneered by the Whipple group who made the first unambiguous detection of TeV $\gamma$-rays from the Crab Nebula in the year $1989$  
\cite{crab}. Subsequently, over the next decade, $\gamma$-rays  were detected from various astrophysical sources \cite{review20gevtev}. 
This field has seen remarkable progress in recent years with the source count increasing from just a single source in 1989 to 175  
confirmed TeV $\gamma$-ray sources $^*$\footnotetext{\small $*$ http://tevcat.uchicago.edu/} as of now.
Imaging Atmospheric Cherenkov Technique (IACT) is based on detecting the Cherenkov radiation generated by 
the cosmic  $\gamma$-rays when they enter the Earth's atmosphere. Presently three major operational IACT based 
telescopes are MAGIC \cite{magic}, HESS \cite{hessI} and VERITAS \cite{veritas}. The MAGIC telescope consists of 
two 17 m diameter telescopes at the Canary island of La Palma. The analysis energy threshold 
of MAGIC is $\sim 80 GeV$. The VERITAS telescope is an array of four 12 m diameter telescopes at southern Arizona, USA. It  
has an analysis energy threshold of $\sim$ 135 GeV \cite{veritas135}. 
HESS telescope, situated in Namibia, is a mixed array consisting of four 12 m telescopes, named as HESS-I 
and one 28 m large size telescope, named as HESS-II. HESS-I alone operates at an 
analysis energy threshold of $\sim$ 158 GeV \cite{hessPC} 
whereas the preliminary simulation studies show that HESS-II has an 
analysis energy threshold of $\sim$ 50 GeV \cite{hess50gev}. 
In order to augment the capability of IACT based telescopes in few GeV to few 
TeV energy range, an international consortium of 
worldwide researchers are setting up an open observatory known as the Cherenkov 
Telescope Array (CTA) \cite{expacta2011}. CTA will consist of two large arrays 
of IACT based telescopes, one in the Northern Hemisphere with 
an emphasis to study  extragalactic objects and a second array in the Southern 
Hemisphere to concentrate on galactic sources. The Southern array which is being 
set up first will deploy telescopes of various diameters 
to cater to the wide energy range of few tens of GeV to 
few tens of TeV. A compact array 
of 4 x 23 m diameter telescopes will cater to the lower end of energy range. 

%

In the same endeavour, a new Indian initiative in gamma-ray astronomy,
the Himalayan Gamma Ray Observatory (HIGRO), is setting up an IACT based telescope known as the {\bf MACE} ({\bf M}ajor 
{\bf A}tmospheric {\bf C}herenkov {\bf E}xperiment) at Hanle in the Ladakh region of northern India. MACE, 
presently being set up at an altitude of 4270m, is a 21m diameter 
telescope with a total light collector area of $\sim 337$ $m^2$ with effective focal length of $\sim 25$ $m^2$. 
Compared to the high altitude of MACE telescope, MAGIC, HESS and VERITAS telescopes 
are operational at an altitude of 2225 m, 1800 m and 1275 m respectively. 
The idea of an IACT based telescope at high altitude (5 km) was introduced by 
Aharonian et al. (2006) \cite{konopelko}. They discussed the 
concept of a stereoscopic array of several large imaging atmospheric Cherenkov telescopes having 
an  energy threshold of 5 GeV. Although it should be noted that they discussed about   
a stereo array, whereas MACE is a standalone single telescope. The stereoscopic approach has many 
advantages compared to the stand alone IACT. This approach allows unambiguous reconstruction of shower 
parameters. It also leads to effective suppression of night sky background and muon background 
because of the reduction in  the random coincidences, leading to reduced pixel trigger threshold 
and hence lower energy threshold. In addition to it, the hadronic showers are rejected more 
efficiently compared to a single IACT based telescope on shape cuts in multiple views. The simultaneous 
observation of air shower by stereoscopic telescope, compared to a single telescope, leads to 
improved shower direction reconstruction as well as core location. The main advantage of 
having a stereoscopic array of 20 m diameter telescope at an altitude of 5 km 
is very low $\gamma$ ray energy threshold $\sim 5$ GeV on account of 
less absorption of Cherenkov photon, as well due to geometric effect on account of higher 
altitude leading to higher photon density. 

In the present work, we will discuss the preliminary sensitivity estimate of the MACE telescope. 
This study is organized as follows. In section \ref{intro}, we will introduce the MACE telescope. In  section \ref{generation} , we will discuss the generation of 
Monte Carlo simulation database along with the technique used in the field of IACT. Section \ref{rforest} will discuss the Random Forest method.
In section 5, we will define the Integral sensitivity and its estimation. Results and discussion will be presented in section 6. 
Finally we will conclude the study along with the planned future studies.

\section{The MACE Telescope}
\label{intro}
\begin{figure}[!h]
\begin{center}
\includegraphics[width=0.75\textwidth, angle =0, clip]{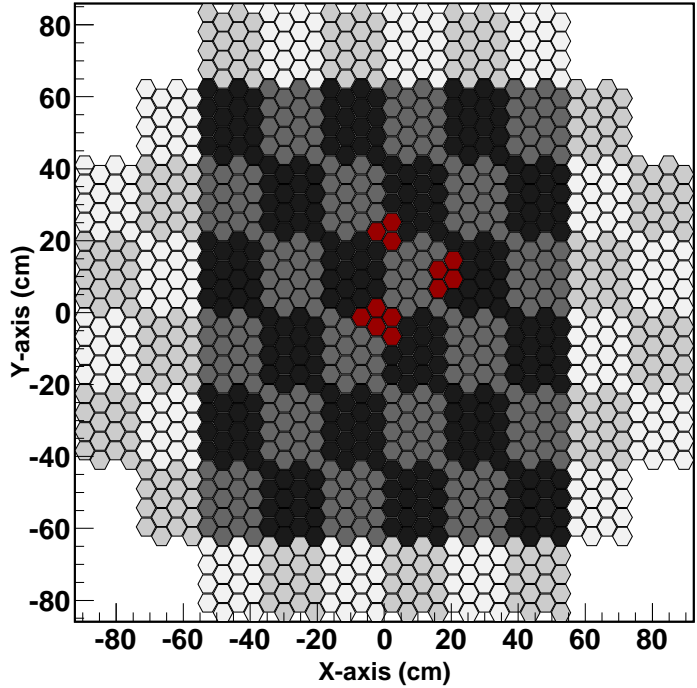}
\caption{\label{Figure.} The MACE camera layout showing a total of 1088 pixels divided into 68 modules with 16 PMT each. The trigger is generated from the darker shaded area  
consisting of 576 pixels. Typical trigger pattern of various programmable scheme (3,4,5 CNNN) is shown in red color.}\label{Figure:maceCam}
\end{center}
\end{figure}

The MACE is an Indian effort to set up a very high energy (VHE) $\gamma$-ray IACT based telescope. The MACE telescope 
is presently being installed at Hanle in Ladakh, India (32$^0$  46$^{'}$ 46$^{"}$ N, 78$^0$ 58$^{'}$ 35$^{"}$ E) at an altitude of 4270 m a.s.l. It is a 21m diameter 
telescope which will deploy a photomultiplier tube (PMT) based imaging camera consisting of 1088 pixels. 
The diameter of each PMT is 38 mm with angular resolution of $0.125^{\circ}$ and  an optical field of view 
of $\sim 3.4^{\circ} \times 4^{\circ}$. In order to reduce the dead space between the PMTs, a light concentrator having a hexagonal entry
aperture of 55 mm  and a circular exit aperture of 32 mm is placed on top of the PMTs. 
The imaging camera of MACE telescope has been designed in a modular manner consisting of 68 modules of 16 channels each. The MACE camera layout has been shown 
in Figure \ref{Figure:maceCam}. The detailed trigger scheme is described in \cite{chinmayB}. Out of 1088 pixels, the innermost 576 pixels ($24 \times 24$) will be used for 
trigger generation. The trigger field of view is $\sim 2.6^{\circ} \times 3^{\circ}$. Conventionally, the event trigger in an Atmospheric Cherekov Telescope is generated by 
demanding a fast coincidence between few PMTs, generally 2-6. Since the $\gamma$-ray images are more compact compared to the background events, the chance coincidence 
rate can be reduced by limiting the n-fold topological combination of pixels. The MACE telescope uses nearest neighbour close cluster 
patterns. It has an option of using various programmable trigger configuration such as 2 CCNN (close cluster nearest neighbour pairs), 3 CCNN (close cluster nearest 
neighbour triplets), 4 CCNN (close cluster nearest neighbour quadruplets). Here CCNN is defined as the pattern in which if any
one of the fired pixels is removed, the remaining pixels should be still adjacent. MACE trigger will be generated in two stages: In the first stage a m-fold 
coincidence in a module between 
nearest neighbor pixels with a coincidence gate width of ∼ 5 ns is demanded . In the second stage, partial triggers from individual modules are collated so that events, spread
over adjacent modules (2,3 or 4) satisfying the chosen multiplicity condition generate the trigger.  
The MACE camera layout with 3, 4 and 5 CCNN trigger patterns is shown in Figure \ref{Figure:maceCam}. 

MACE Telescope will use a quasi-parabolic light collector to image the Cerenkov flash on a PMT camera. The light collector of the telescope will be made of 356 mirror panels of 
984 mm $\times$ 984 mm size fixed at a square pitch of 1008 mm on a paraboloid shape  dish. Each panel will consists of four 488 mm $\times$ 488 mm facets of spherical 
mirrors made of aluminum with a $SiO_2$ coating. The total light collector area will be $\sim 337 m^2$. The mirror facets have a graded focal length of 25-26.5 m 
which ensures that the on-axis spot size is minimum at the focal plane.

The MACE is an Indian effort to set up a very high energy (VHE) $\gamma$-ray IACT based telescope. The installation of this telescope
is presently going on at Hanle in Ladakh, India (32$^0$  46$^{'}$ 46$^{"}$ N, 78$^0$ 58$^{'}$ 35$^{"}$ E) at an altitude of 4270 m a.s.l. It is a 21m diameter 
telescope which will deploy a photomultiplier tube (PMT) based imaging camera consisting of 1088 pixels. 
The diameter of each PMT is 38 mm with angular resolution of $0.125^{\circ}$ and  an optical field of view 
of $\sim 3.4^{\circ} \times 4^{\circ}$. In order to reduce the dead space between the PMTs, a light concentrator having a hexagonal entry
aperture of 55 mm  and a circular exit aperture of 32 mm is placed on top of the PMTs. 
The imaging camera of MACE telescope has been designed in a modular manner consisting of 68 modules of 16 channels each. The MACE camera layout has been shown 
in Figure \ref{Figure:maceCam}. The detailed trigger scheme is described in \cite{chinmayB}. Out of 1088 pixels, the innermost 576 pixels ($24 \times 24$) will be used for 
trigger generation. The trigger field of view is $\sim 2.6^{\circ} \times 3^{\circ}$. Conventionally, the event trigger in an Atmospheric Cherenkov Telescope is generated by 
demanding a fast coincidence between few PMTs, generally 2-6. Since the $\gamma$-ray images are more compact compared to the background events, the chance coincidence 
rate can be reduced by limiting the n-fold topological combination of pixels. The MACE telescope uses nearest neighbour close cluster 
patterns. It has an option of using various programmable trigger configuration such as 2 CCNN (close cluster nearest neighbour pairs), 3 CCNN (close cluster nearest 
neighbour triplets), 4 CCNN (close cluster nearest neighbour quadruplets). Here CCNN is defined as the pattern in which if any
one of the fired pixels is removed, the remaining pixels should be still adjacent. MACE trigger will be generated in two stages: In the first stage a m-fold coincidence in a module between 
nearest neighbor pixels with a coincidence gate width of ∼ 5 ns is demanded . In the second stage, partial triggers from individual modules are collated so that events, spread
over adjacent modules (2,3 or 4) satisfying the chosen multiplicity condition generate the trigger.  
The MACE camera layout with 3, 4 and 5 CCNN trigger patterns is shown in Figure \ref{Figure:maceCam}. MACE Telescope will 
use a quasi-parabolic light collector to image the Cerenkov flash on a PMT camera. The light collector of the telescope will be made of 356 mirror panels of 
984 mm $\times$ 984 mm size fixed at a square pitch of 1008 mm on a paraboloid shape  dish. Each panel will consists of four 488 mm $\times$ 488 mm facets of spherical 
mirrors made of aluminum with a $SiO_2$ coating. The total light collector area will be $\sim 337 m^2$. The mirror facets have a graded focal length of 25-26.5 m 
which ensures that the on-axis spot size is minimum at the focal plane.

\section{Generation of Monte Carlo simulation database}
\label{generation}
A Monte Carlo simulation database at 5$^\circ$ zenith angle was generated. The details are as follows:
\subsection{Simulation Database}
The extensive air shower (EAS) library for MACE simulation was generated using a standard air shower simulation
package CORSIKA(v6.990). This code is developed at Karlsruhe university and available for use on request. This Monte
Carlo method based code simulates the secondary particle cascade generation in the atmosphere as high energy primary
particle enters the Earth's atmosphere and undergoes many electromagnetic as well as hadronic interactions. Many models
to simulate these interactions are available in this code and the Cerenkov light generation by the relativistic 
leptons as they zip through the atmosphere is also simulated. For this work, we have used EGS4 model for the electromagnetic 
interactions. For low energy (upto \~100GeV) hadronic interactions, we have used GHEISHA model while high energy interactions
are modelled using QGSJet-I model. We have used CEFFIC option to incorporate the en-route absorption of Cerenkov photon. 
We have used US standard atmosphere model supplied with CORSIKA package. The observation level was set to the altitude of 
the telescope site. The Earth's magnetic field values measured at the Hanle 
are 31.95 $\mu T$ for horizontal component along local north direction and 38.49 $\mu T$ for vertical 
component in downwards direction. The Cerenkov photons were stored within the wavelength range 240$nm$ -- 650$nm$ as
the MACE camera PMTs are sensitive in this range. 

Using this setup, we generated EAS library for four primary particles -- $\gamma$-ray, proton, electron and $\alpha$ particle.
The spectral indices used are 2.59 \cite{crabHEGRA} for $\gamma$-ray, 2.7 for proton, 3.26 for electron and 2.63 for $\alpha$ particle.
Other input parameters which vary with particle type, are listed in Table \ref{Table:database}. In this work we focused on
estimating the sensitivity of the telescope for on-axis point sources only (view cone angle is maintained at zero value
for $\gamma$-ray showers). The protons, electrons and alpha particles are simulated diffusively in a certain view cone angles, as 
described in the Table \ref{Table:database}.   
It evident from this table that we generated more than 34 million $\gamma$-ray 
showers and nearly 1 billion showers for cosmic rays in two energy band encompassing over 3 decades of energies using
three high-end workstations comprising of 32 processors each during the actual runtime of more than six months.

The telescope simulation program took these shower events as input and ray-traced them to the focal plane of the telescope.
This simulation used the same optical design of the telescope including the pixel layout in the camera as described in section 2.
The two level trigger logic used in simulation exactly mimics the actual trigger design to be implemented in the MACE telescope. 
For triggering criteria \cite{chinmayB}, we have used 9 p.e. as the discrimination threshold for a pixel and the trigger configuration of 4 CCNN pixels was used.

\noindent
\begin{table}[htbp]
	\renewcommand{\arraystretch}{0.6}
	\begin{center}
	\begin{tabular}{|c|c|c|c|c|c|}
		\hline
		particle & Energy Range &  View cone & Scatter    & No. of showers & No. of Triggered  \\
		type     &              &  angle(deg) & radius(m) &  (million)     & showers \\
		\hline
		\multirow{2}{*}{$\gamma$-ray}
			& 10GeV--20TeV &  $0^o$ & 400 & 22.8 & 1,314,306 \\
		             & 400GeV--20TeV &  $0^o$ & 500 & 12.8  & 1,569,866\\
		\hline
		\multirow{2}{*}{proton}
		    & 20GeV--20TeV &  $3^o$ & 500 & 356 & 308,099 \\
		             & 400GeV--20TeV & $3^o$ & 550 & 64 & 169,221\\
		\hline
		\multirow{2}{*}{electron}
		    & 10GeV--20TeV &  $3^o$ & 500 & 384 & 432,631 \\
		             & 400GeV--20TeV & $3^o$ & 550 & 64 & 330,909\\
		\hline
		    $\alpha$ particle & 100GeV--20TeV &  $3.5^o$ & 550 & 326 & 391,653 \\
		\hline
	\end{tabular}
	\end{center}
	\caption{Monte Carlo simulation database}
	\label{Table:database}
\end{table}

\begin{figure}[!h]
\begin{center}
\includegraphics[totalheight=3.5cm]{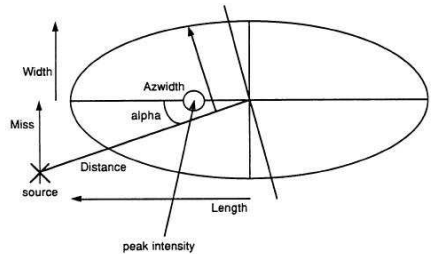}
\caption{{\it schematic presentation of Hillas (image) parameters}}\label{Figure:image}
\label{fig_had}
\end{center}
\end{figure}

For simulation studies, a general 
purpose IACT simulation code was employed \cite{chinmayB} to asses the performance of the MACE telescope. In brief, the general purpose simulation code incorporated 
the specification for the MACE telescope like camera, reflector and trigger configuration along with the wavelength dependent photon absorption. A poissonian 
Night Sky Background (NSB) background in each pixel was added in this code. The NSB observation was carried out at the MACE telescope installation site \cite{stalin}
during 2003 - 2007 using CCD images taken with the Himalayan Chandra Telescope (HCT). The estimated NSB rate per pixel turns out to be $\sim 1.46$ photoelectrons \cite{chinmayB}.
The Cherenkon photons falling on the image plane of the camera underwent usual image analysis procedure developed by Hillas \cite{Hillas}. 
The spatial distribution (of Cherenkov photons)  generates different patterns for signal ($\gamma$-rays generated Cherenkov photons) as well as for background (Protons, 
Electrons, Alpha generated Cherenkov photons) events. This difference in shape and size is exploited for the extraction of signal. 
A typical schematic representation of Hillas parameters is shown in 
Figure \ref{Figure:image}.
Conventional practice of image cleaning \cite{cleaning} was employed. We used picture and boundary thresholds of 10$\sigma$ and 5$\sigma$ respectively. 
The clean Cerenkov images were characterized by various Hillas image parameters \cite{Hillas} like {\it Length, Width, Distance, Size, Frac2 and alpha}.
The rms spread of Cherenkov light along the major/minor axis of image is known as the {\it Length/Width} of an image. 
A first level measure of $\gamma$-initiated extensive air showers core distance is given by the Distance parameter. 
It represents the distance from the image centroid to the position of $\gamma$-ray source in the 
field of view. The compactness parameter is known as Frac2. It is defined as the ratio of the sum of the two highest 
pixel signal to the sum of all the signals. The total number of photoelectrons in a Cherenkov image is represented by 
the size parameter. The parameter {\it alpha} is the angle of the image between a line joining the centroid of the 
image to the centre of the field of view. It is a measure of the orientation of the shower axis. The $\gamma$-ray 
initiated showers are pointed towards the centre of the imaging camera and hence have very small value of {\it alpha}
compared to the angles subtended by the background events on account of their isotropic distribution. Asymmetry 
parameter is defined as  the third moment of the intensity distribution along the major axis. It describes the 
skew of the image along its major axis. 
\begin{figure}[!h]
\begin{center}
\includegraphics[width=0.25\textwidth, angle =270, clip]{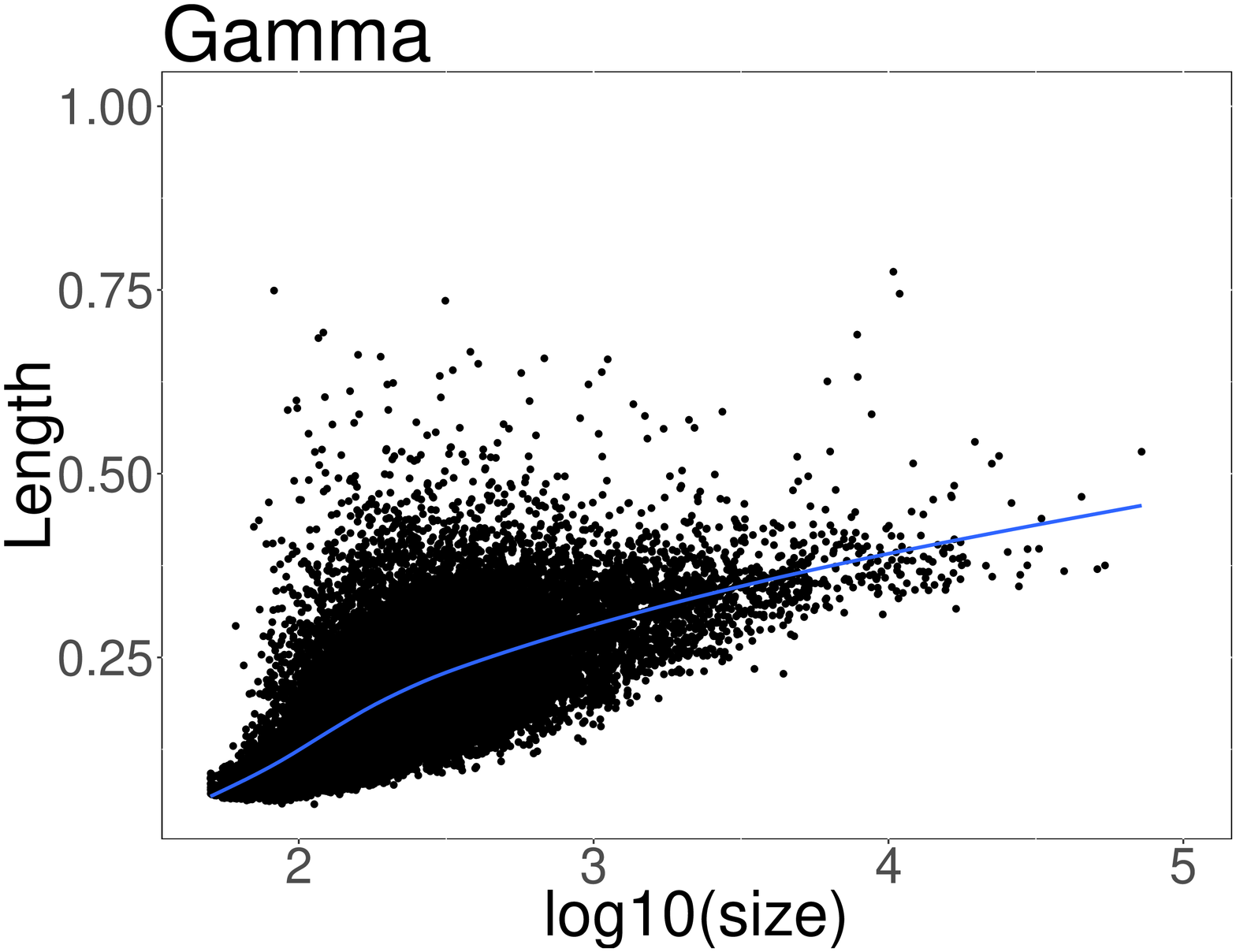}\includegraphics[width=0.25\textwidth, angle =270, clip]{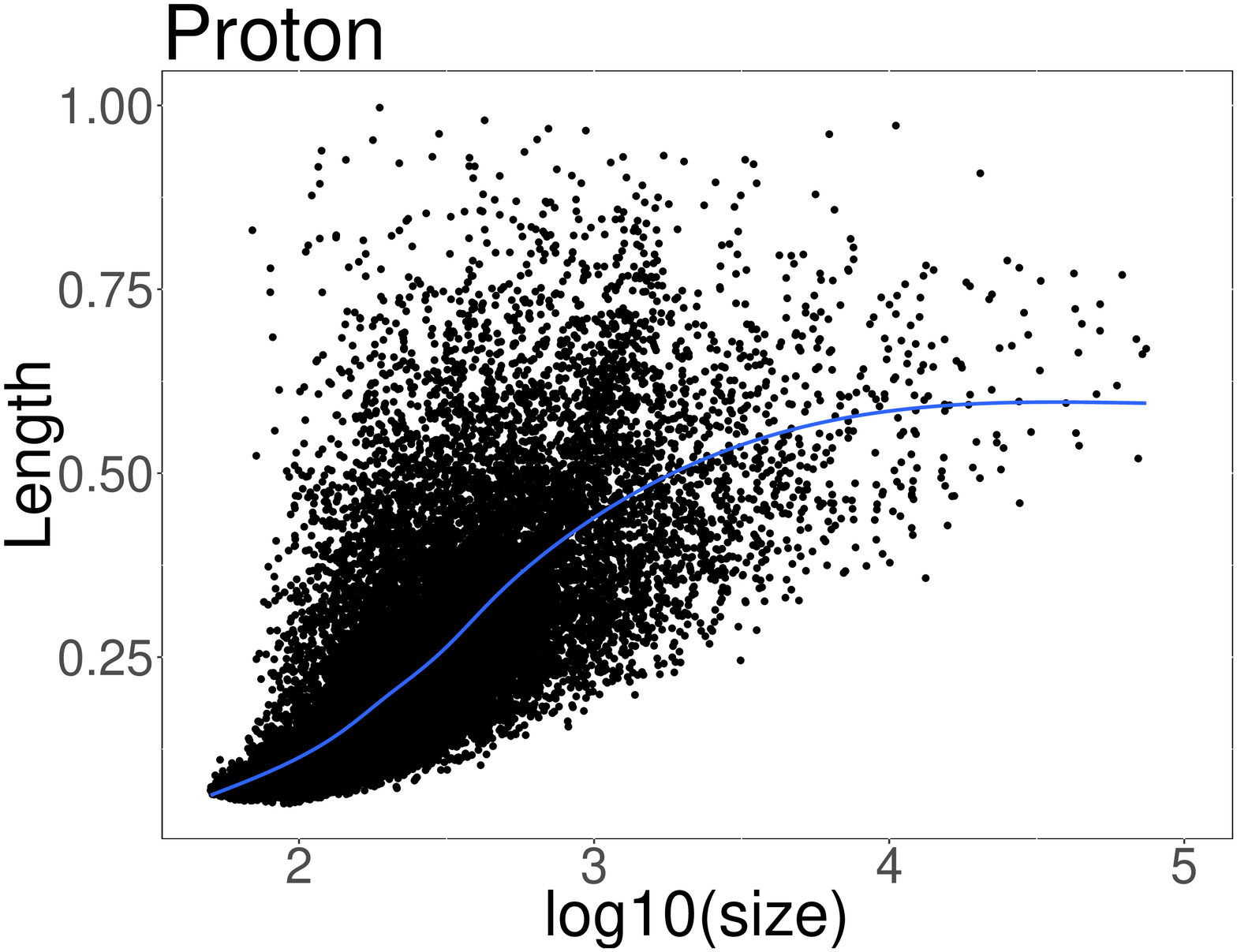}
\includegraphics[width=0.25\textwidth, angle =270, clip]{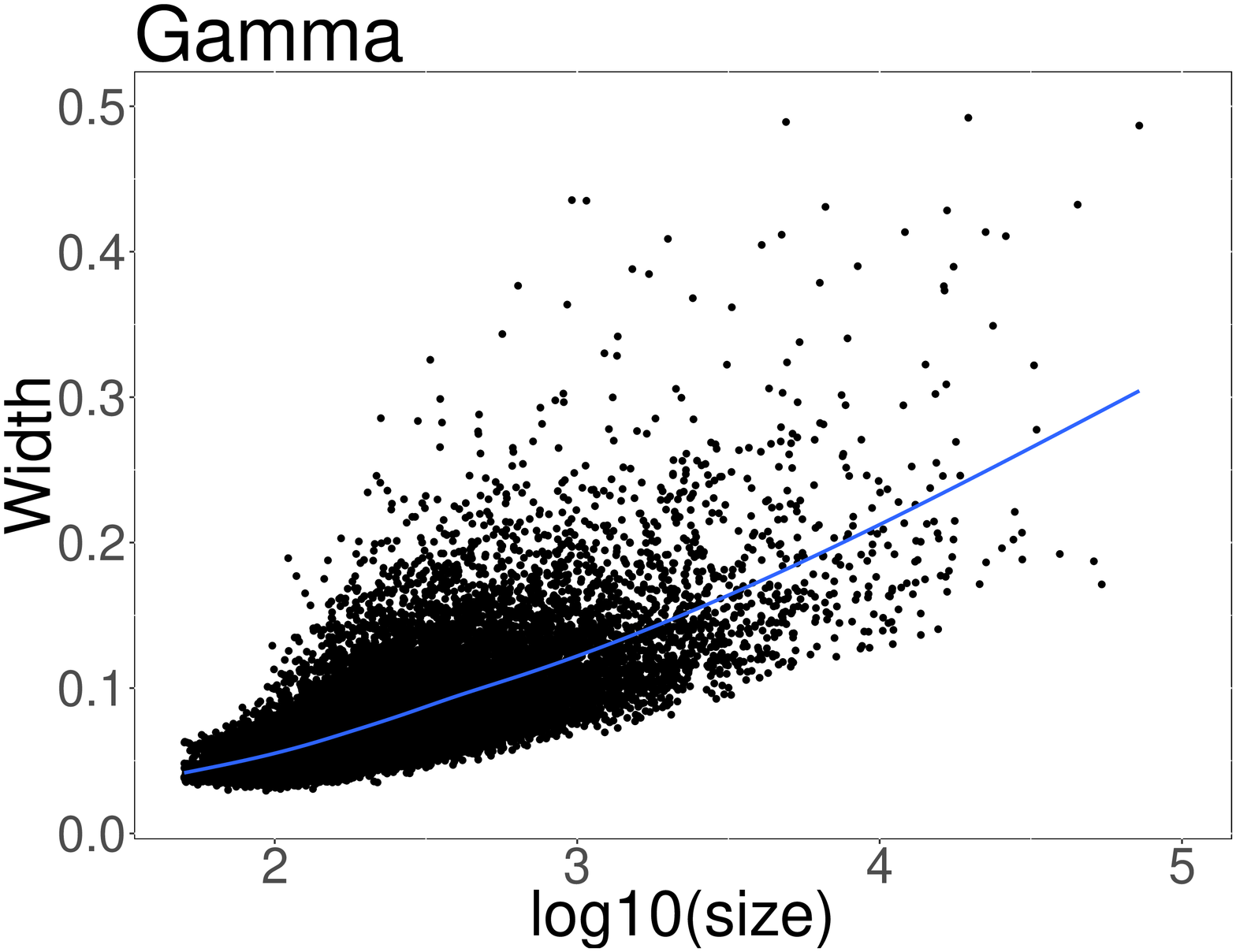}\includegraphics[width=0.25\textwidth, angle =270, clip]{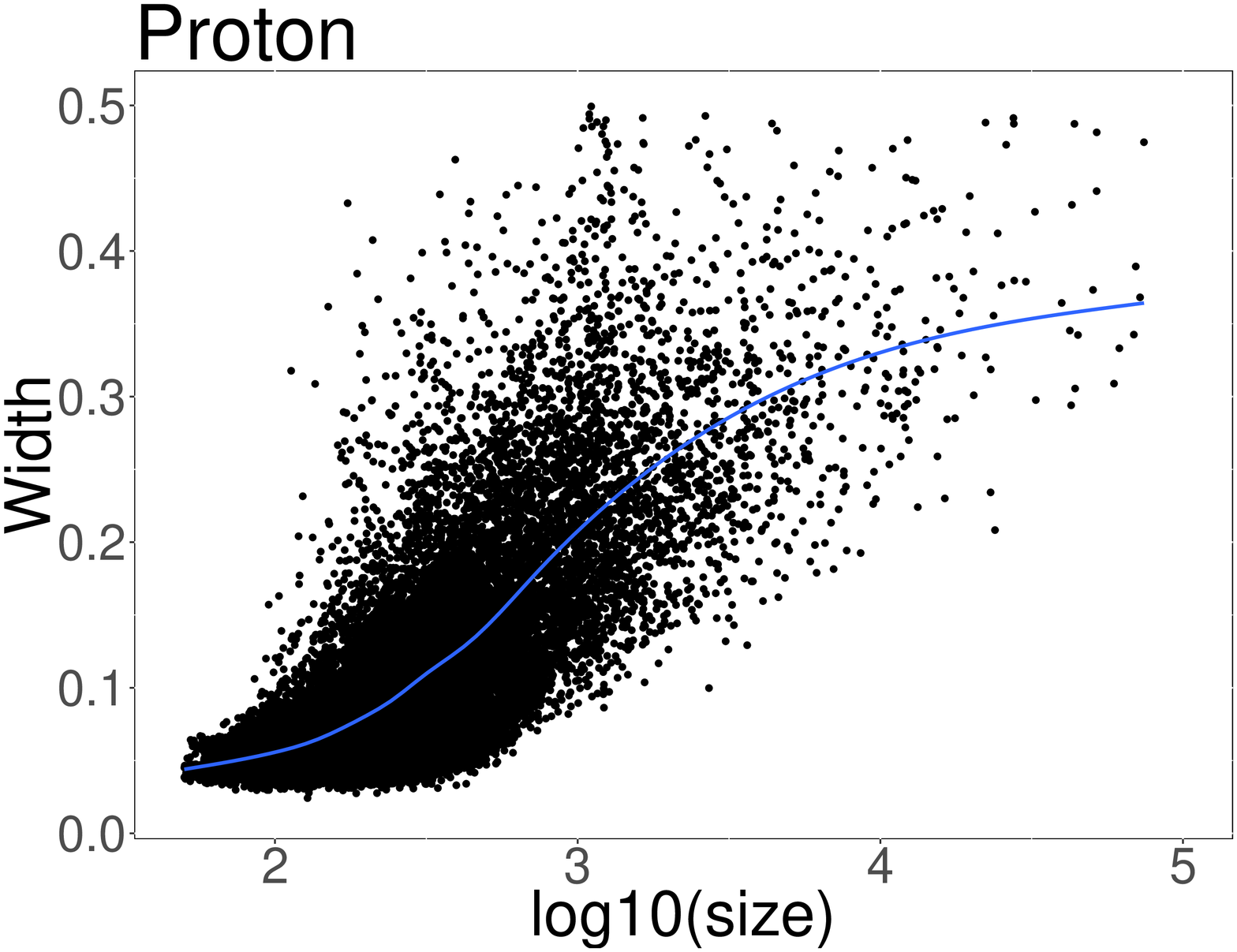}
\caption{\label{Figure.}Length and Width distribution of Monte Carlo Gamma and Protons. The profile plot exhibits the mean value of corresponding Hillas parameters.}\label{Figure:noele}
\end{center}
\end{figure}
In the present work, we have used Length, Width, Distance, Frac2, 
Size and Asymmetry Hillas parameters. In addition to these parameters, we have 
used an additional parameter, known as the {\it leakage} parameter. It is defined as the 
ratio between the light content in the camera two outer most pixel rings to the total light content.
Leakage is used as a precut parameter. We have applied a leakage cut of 10\% in this study. This cut  
removes events showing a concentration of more than 10 \% of the image size in the two outermost pixel rings. 
If such events are not rejected, it lead to a general distortion of the image parameters. A typical profile plot for 
{\it Length} and {\it Width} parameters with {\it size} for Monte Carlo gamma and protons is shown in Figure 
\ref{Figure:noele}. 
We have also shown the Length and width distribution of $\gamma$ and protons with 
Energy is shown in Figure \ref{Figure:lwedist}. This Figure is shown in the Appendix.

\section{Random Forest}
\label{rforest}

Random Forest (RF) is one of the many flexible multivariate machine learning methods. The algorithm for 
RF was developed by Leo Breiman and Adele Cutler$^*$ \footnotetext{\small $*$ http://www.stat.berkeley.edu/∼breiman/RandomForests/} 
and can be used  for classification and regression problems. Random Forest is an ensemble of simple tree predictors where each tree 
makes a prediction and final prediction is made by aggregating over the ensemble. A nice description of Random Forest methods is given in 
\cite{mradul1,magicRF}. The classification tree forms the basic building block of Random Forest method. It is 
constructed by binary recursively partitioning the data set. Each partitioning splits the data sets into different branches. 
In the present work, the basic aim of the application of Random Forest is to segregate Cherenkov photons generated by the 
cosmic $\gamma$-rays from the background consisting of the Cherenkov photons generated from other cosmic ray particles like 
protons, electrons and alpha particles. For this purpose, the present problem is treated as a binary classification problem 
where one of the classes $\gamma$ is to be segregated from the other class consisting of protons, electrons and alpha particles. 
Each entry of training sample consisting of two classes are known as \textit{events}. Each event is characterized by a vector 
containing various image (Hillas) parameters defined earlier.

\begin{figure}[!h]
\begin{center}
\includegraphics[height= 6cm, width =5cm, angle = -90]{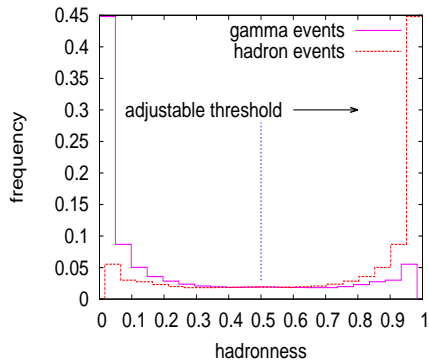}
\caption{\label{Figure.} The mean hadronness for the two test samples of gammas and hadrons.}\label{Figure:hadronness}
\end{center}
\end{figure}

The cuts on the dataset are generated by taking 
the various components of vector having various Hillas parameter. 
In the present work, we used a simulation database with details shown in Table \ref{Table:database}. 
We used 35000 events of each species for the training purpose. Rest of the triggered events were used as a test sample. 
In this study, we have used the original RF code implemented in Fortran \footnote{\small http://www.stat.berkeley.edu/∼breiman/RandomForests}. 
One of the strengths of RF is that there are only two parameters which are to be tuned to minimize the prediction error. One parameter is known as 
\textit{mtry}, i.e. number of input variables chosen randomly at each split and the second parameter is \textit{ntree}, i.e. the number of trees 
to be generated.  There is another parameter, known as \textit{nodesize} which decides the minimum number of data points  after which the binary 
partitioning is stopped. We generated a total of 500 trees each and  used mtry = 2,3. Both the values of mtry gave similar results. The parameter 
\textit{nodesize} was kept as 10. An event is classified as $\gamma$-like or non $\gamma$-like on the basis of hadronness parameter. The hadronness 
of two samples of gamma and hadrons is shown in Figure \ref{Figure:hadronness}.

\section{Integral Sensitivity}
\begin{figure}[!h]
\begin{center}
\includegraphics[height=4.2cm, width=4.6cm]{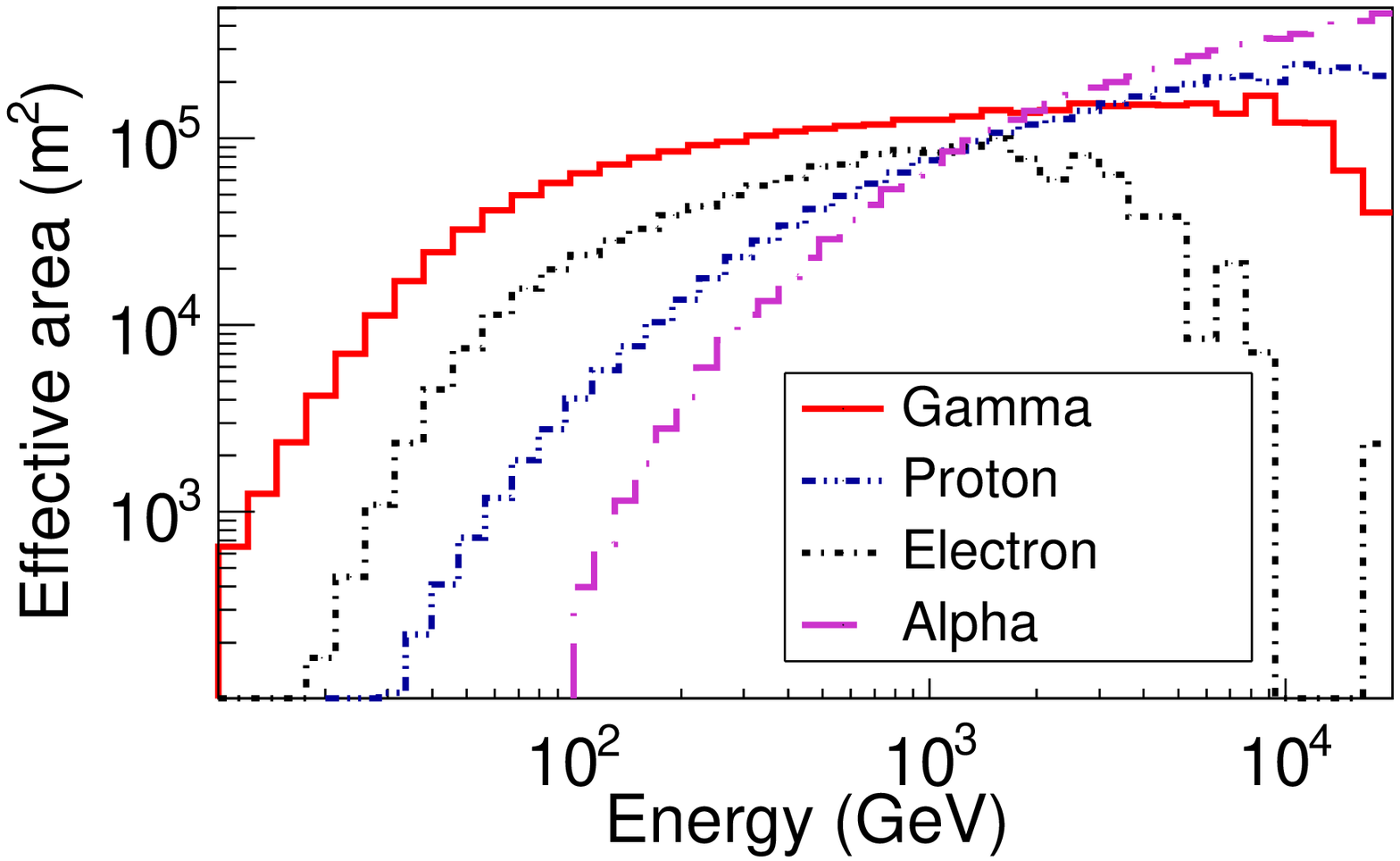}\includegraphics[height=4.0cm, width=4.5cm,angle=0,clip]{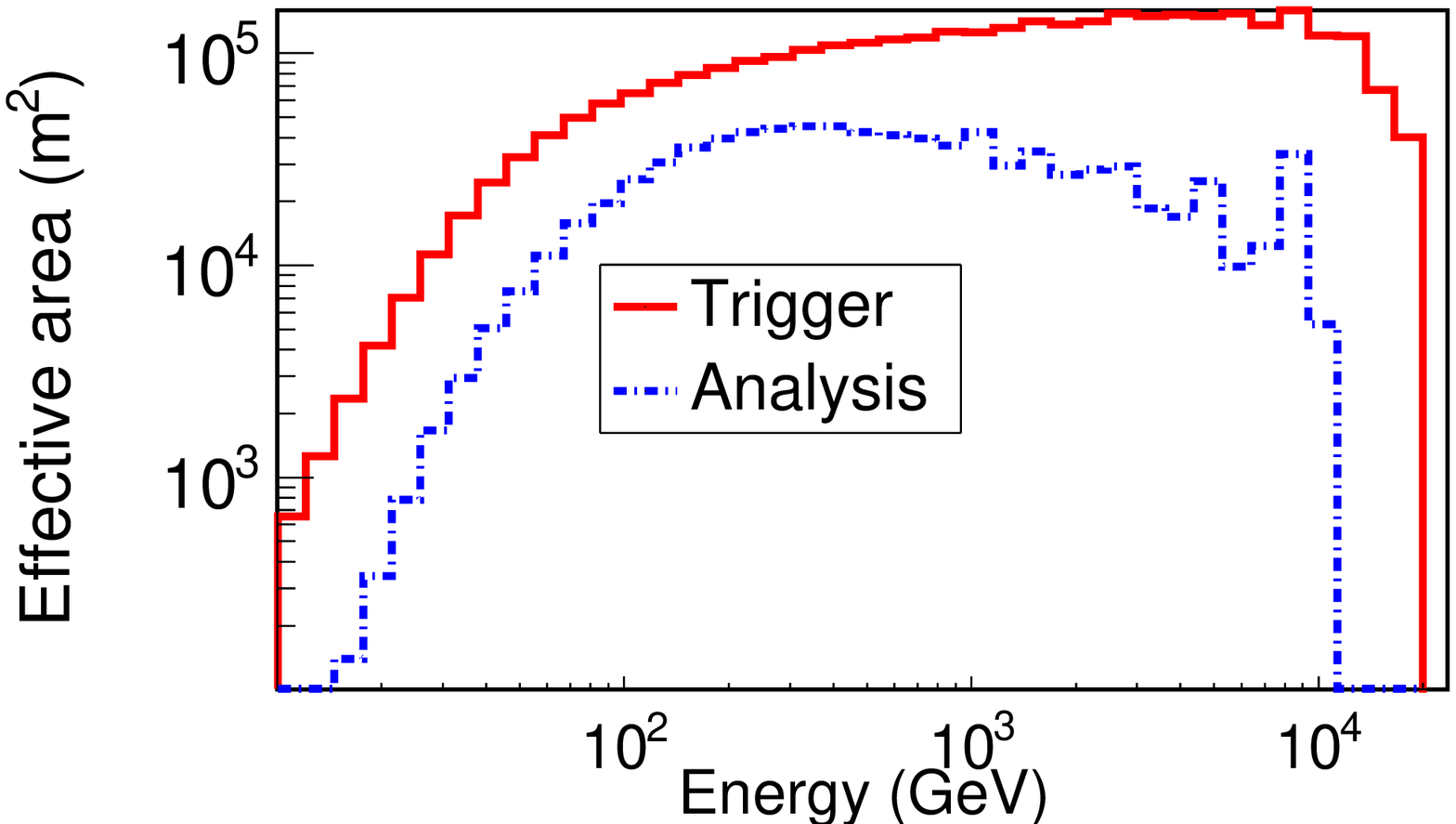}
\caption{\label{Figure.}Left panel shows the Trigger effective area for the MACE telescope for various cosmic ray particles like Electrons, Alpha, proton as well as cosmic $\gamma$-rays. The right panel
shows the trigger and analysis effective area for cosmic $\gamma$-rays.}\label{Figure:EA}
\end{center}
\end{figure}

The sensitivity of a gamma-ray telescope is determined by its ability to detect
gamma-ray signal in overwhelming presence of cosmic ray background. 
The sensitivity of a telescope depends on the area over which the detector is sensitive 
to the events. This area is known as the effective area of the telescope. For an IACT based telescope 
this effective area is orders of magnitude bigger than the size of the light collector. The  
effective area of MACE telescope for the $\gamma$-rays originating from an on-axis point source
is estimated by integrating the total retention factor for $\gamma$-rays over all the core 
distances from the telescope, as shown in the equation \ref{eq:g_eff}. Normally, the effective area is 
also a function of zenith angle but since we are reporting the results at a particular zenith angle, the zenith angle
dependence is not shown explicitly. 
\begin{equation}\label{effagamma}
	A_{eff}^{\gamma}(E) = 2\pi \int\limits_0^\infty dR \, R \times \eta_{total}^\gamma(E,R)
\end{equation} \label{eq:g_eff}
Here, $\eta_{total}^\gamma$ is the total retention factor for $\gamma$-rays and it
includes the trigger retention factor as well as the retention factor due to $\gamma$-selection cut. 
If $dN_{simulated}$ is the number of events used in the simulation and $dN_{selected}(E,r)$ 
is the number of events triggered the telescope as well as selected on the basis of 
$\gamma$-selection cut as described above then the total retention factor is defined as,
\begin{equation}
	\eta_{total}^\gamma(E,r) = \eta_{trigger}^\gamma(E,r)\cdot\eta_{cut}^\gamma(E,r) 
	                         = \frac{dN_{selected} (E,r)}{dN_{simulated} (E,r)}
\end{equation}
Similarly the effective area for each of the cosmic ray particles (protons, electron
and $\alpha$ particles) is estimated by integrating its respective total retention factor
at different core distances over all the core distances as well as over all the angles within 
the view-cone solid angle  around the optic axis of the telescope as shown in the following equation;
\begin{equation}
	A_{eff}^{\mathbb{X}}(E) = 2\pi \int\limits_0^\infty dR \, R 
	\int\limits_0^{\Omega_{vc}} d\Omega \; \eta_{total}^{\mathbb{X}}(E,R,\Omega)\; ; 
	\quad\quad ({\mathbb{X}} = proton,\; electron,\; \alpha)
\end{equation}
And the total retention factor for cosmic ray particles is determined in same manner as in 
the case of $\gamma$-rays and is shown below,
\begin{equation}
	\eta_{total}^{\mathbb{X}}(E,r,\Omega) = \frac{dN_{selected} (E,r,\Omega)}{dN_{simulated} (E,r,\Omega)}\; ;
	\quad\quad ({\mathbb{X}} = proton,\; electron,\; \alpha)
\end{equation}
The trigger effective areas for $\gamma$-rays and all the three cosmic ray species are shown in 
the left panel of Figure \ref{Figure:EA}. The after-analysis effective area, $A_{eff}^\gamma(E)$,
for $\gamma$-rays is shown in the right panel of Figure \ref{Figure:EA} along with the trigger
effective area for direct comparison. We also estimated the ``analysis effective area'' by 
applying the hadronness cut value of 0.9 
with size $>$ 50 photoelectrons. The value of hadronness is chosen by optimizing the sensitivity.  
It is clear from the right panel of Figure \ref{Figure:EA} that the analysis effective area for $\gamma$ rays for the MACE telescope 
is much lower than the trigger effective area. This effect is especially visible 
for higher energies. The effective area above  1 TeV shows large fluctuations on account of poor shower statistics. 

\noindent 
The integral sensitivity of an IACT telescope is determined by calculating the 
minimum detectable integral flux of gamma-rays from Crab-like source having energies more 
than pre-determined threshold energy at $5\sigma$ confidence level in 50 hours of observation.
The threshold energy used in this estimation is the energy at which the differential  
rate of gamma-ray events having sizes more than certain pre-determined size is maximum.  
For integral sensitivity estimation for MACE telescope, we have used 50 p.e. as minimum size and 
subsequently the size is increased by 1.5 times the earlier value. For each size value we 
demanded that the source flux for energies more than the threshold energy should be such 
that the significance as defined by equation (\ref{sigeq}) had to be $5\sigma$ in 50 hours of observation time 
and also at least 10 $\gamma$-rays should be selected.
\begin{equation}\label{sigeq}
	\mathbb{N}_\sigma = \frac{N_\gamma}{\sqrt{N_\gamma+2N_{CR}}} 
	\quad\quad \mathrm{where,}\; N_{CR} = N_p+N_e+N_\alpha
\end{equation}
Here, $N_\gamma$ is the number of $\gamma$-rays selected after the analysis as described in 
the earlier section and it is defined by the following equation,
\begin{equation}
	N_\gamma = T_{obs} \int\limits_{E_{th}}^{E_{max}} dE \;\, \frac{dN_\gamma (E)}{dE} \, A_{eff}^\gamma(E)
\end{equation}
We have used Crab Nebula spectrum as measured by HEGRA \cite{crabHEGRA} for this calculation.
And the effective area is calculated using equation (\ref{effagamma}). 
Similarly the number of cosmic ray particles selected after the analysis are estimated using
the following equation,
\begin{equation}
	N_{\mathbb{X}} = T_{obs} \int\limits^{\Omega_{vc}} d\Omega
	\int\limits_{E_{th}}^{E_{max}} dE \;\, 
	\frac{dN_{\mathbb{X}}(E,\Omega)}{dE} \, A_{eff}^\gamma(E) 
	\quad ({\mathbb{X}} = proton,\; electron,\; \alpha)
\end{equation}
Here, $\frac{dN_{\mathbb{X}}(E,\Omega)}{dE}$ is the cosmic rays spectrum for protons, 
electrons and alpha particles  as reported in Saino et al. (2004) \cite{spectraCosmic1} and Wiebel-Sooth et al.(1998) \cite{spectraCosmic2}.
The cosmic ray flux is assumed to be isotropic within the  view-cone solid angle. 

\begin{figure}[!h]
\begin{center}
\includegraphics[width=0.5\textwidth,angle=270,clip]{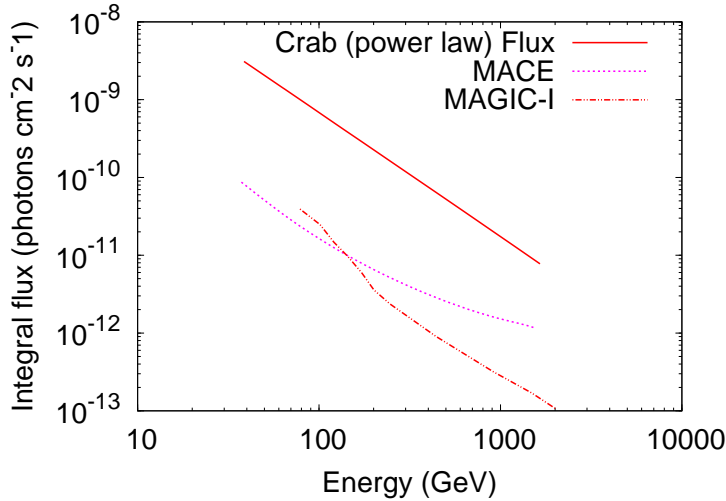}
\caption{\label{Figure.}Integral sensitivity of the MACE telescope at $5^{0}$ zenith angle. The sensitivity of the MAGIC-I telescope is also shown.}\label{Figure:specs} 
\end{center}
\end{figure}

Armed with all the required parameters, here we have estimated the sensitivity curve of the MACE 
telescope. The analysis energy threshold $*$ \footnote{\small The energy 
at which differential rate curve for a particular progenitor particle above a certain size range 
peaks, is called the  \textit{Threshold energy} of the telescope for that particle} for the MACE telescope is  38 GeV corresponding to 
50 p.e. but it does not offer the best sensitivity. 
The integral sensitivity for point like sources with Crab Nebula-like spectrum 
above 38 GeV is $\sim 2.7\%$ of Crab Nebula flux at 5 $\sigma$ statistical 
significance level in 50 hrs of observation. The integral sensitivity of the MACE telescope is shown in Figure \ref{Figure:specs}. 
Along with the MACE sensitivity, we have also shown the sensitivity of the MAGIC-I telescope \cite{magic1}. 
It is clear from Figure \ref{Figure:specs} that compared to the MAGIC-I telescope, MACE telescope will have a lower energy threshold (as expected on account of
higher altitude). Also, it is clear that MACE telescope will be more sensitive than the MAGIC-I telescope upto 150 GeV.
 
\section{Results and Discussion}
We have estimated the sensitivity of the MACE telescope where the $\gamma$-hadron segregation was carried out by 
employing the Random Forest method. The sensitivity was estimated by generating a substantially 
large  Monte Carlo simulation database at $5 ^{\circ}$ zenith angle. We generated $\sim 1 \times 10^9$ EAS for various 
relevant cosmic ray species ($\gamma$, Proton, Electron, alpha particles). The MACE 
telescope is very sensitive in the low energy range, especially below $150$ GeV. The analysis energy threshold of 
MACE telescope turns out to be $\sim 38$ GeV for size cut $>$ 50 photoelectrons. 
It will be able to detect a minimum of $\sim 2.7 \%$ Crab flux in 50 hrs of observation. 
Since MACE telescope in 
the present study will operate in mono mode, it is worthwhile to compare the sensitivity of the MACE telescope with respect to the MAGIC-I telescope. 
Figure \ref{Figure:specs} shows that MACE telescope is able to achieve a better analysis energy threshold energy of $\sim 38$ GeV whereas 
the analysis energy threshold energy of the MAGIC-I telescope is $\sim 80$ GeV. 
Apart from the lower energy threshold, MACE telescope appears to be more sensitive than the MAGIC-I telescope up to an energy of $\sim 150$ GeV. 
The high altitude of MACE telescope compared to the MAGIC-I telescope leads to a lower energy threshold compared to the MAGIC-I telescope. It was demonstrated 
\cite{konopelko} that stereoscopic array of 20 m diameter IACTs based telescope installed at an altitude of  $\sim 5$ km can achieve a $\gamma$ ray threshold energy of 
$\sim 5$ GeV. Therefore, the MACE telescope on account of higher altitude is expected to achieve a low energy threshold.  It is also shown \cite{chinmayB} 
that for a $\sim 1300$ m altitude, the Cherenkov photon density is $\sim 0.5$ $photons/m^2$, while for Hanle altitude it is $\sim 0.9$ $photons/m^2$ 
up to a core distance of $\sim 100$ m for $\gamma$ rays of energy 10 GeV.  The increase in Cherenkov photon density with increasing altitude is more 
pronounced for $\gamma$ ray showers than hadron showers. Therefore the trigger probability 
for $\gamma$ ray showers is more than trigger probability for hadron showers leading to better performance of MACE telescope in the sub 100 GeV energy range.

.

\begin{figure}[htpb]
	\begin{center}
		\includegraphics[height=.6\textwidth, angle=270]{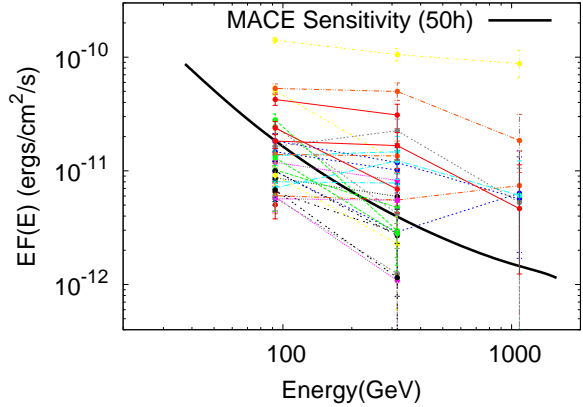}
	\end{center}
	\caption{MACE integral sensitivity curve along with the sources from second high energy Fermi catalogue chosen such that they are detected with high significance 
and are visible at MACE telescope site.}
	\label{fig:senssrc}
\end{figure}

To obtain an immediate source list which can be observed for the estimated sensitivity of MACE telescope,
 we considered the sources listed in second high energy Fermi catalogue \cite{2fhl}. These sources are detected by Fermi
 in the 50 GeV -- 2 TeV energy range over seven years of its operation. To select sources we used the following
 criteria,
\begin{itemize}
	\item the source should be visible from the site of MACE telescope. 
	\item the source should have at least 3$\sigma$ detection in Fermi energy range 171 GeV -- 585 GeV (second energy band
		listed in the catalog). 
	\item the source should have non-zero TS value, signifying mere detection by Fermi, in the Fermi energy range 
		585 GeV -- 2 TeV (third energy band listed 	in the catalog).
\end{itemize}
Above criteria have been set because MACE telescope is most sensitive around 100GeV and sources in Fermi catalog
having 3$\sigma$ detection in 171 GeV -- 585 GeV energy range generally have more than 5$\sigma$  detection in 
50 GeV -- 171 GeV energy range. Therefore those sources are the best choice to study the 
performance of the MACE telescope. The Fermi observed spectra of the selected sources along with the MACE sensitivity 
curve are shown in Figure \ref{fig:senssrc} for comparison. It shows that MACE is expected to detect many more new
sources in future. Table \ref{Table:2fhl} lists the sources selected from the second Fermi high energy catalogue 
visible to the MACE telescope. The list of these sources is given in the appendix.

\section{Conclusion}
We have presented the sensitivity estimate of the MACE telescope. The salient features of the MACE are:
\begin{itemize}
\item The MACE telescope is presently being set up at Hanle, Ladakh, India at an altitude of 4270 meters. Once operational, it will become the highest 
altitude VHE telescope working on the IACT.

\item Due to its high altitude, MACE telescope is able to achieve analysis energy threshold of 38 GeV.

\item The integral sensitivity for point like sources with Crab Nebula-like spectrum 
above 38 GeV is $\sim 2.7\%$ of Crab Nebula flux at 5 $\sigma$ statistical 
significance level in 50 hrs of observation.

\item The sensitivity of the MACE telescope below 150 GeV ($\gamma$-ray energy) is better than the MAGIC-I telescope.

\item From second Fermi high energy catalogue, we selected sources which are visible at Hanle site and are significantly detected
by Fermi/LAT. From Figure \ref{fig:senssrc}, it is evident that MACE will detect these sources.
\end{itemize}

In order to estimate the sensitivity of the MACE telescope in all the zenith angle ranges, we are generating a similar 
Monte Carlo simulation database.It has been shown by the MAGIC group that for a single dish telescope, by using the 
timing information, background gets reduced by a factor two, which in turn results in an enhancement of about a factor 
1.4 of the flux sensitivity to point-like sources \cite{time}. We have also initiated a study to include the timing 
information of the Cherenkov pulses in estimating the revised sensitivity of the MACE telescope. 

\section*{Acknowledgements}
The authors are very thankful to both the anonymous reviewers whose comments/ suggestions lead to a considerable improvement in the final manuscript. We also thank all 
the colleagues of Astrophysical Sciences Division for their inputs and suggestions during the course of this work. The authors are also grateful for the Computer 
Division, BARC for providing excellent computing facility.

\bibliographystyle{h}
\bibliography{mradul-nimpa2}

\begin{thebibliography}{10}

\bibitem{crab}
T.C. {Weekes} et~al.,
\newblock ApJ 342 (1989) 379.

\bibitem{review20gevtev}
M. {de Naurois},
\newblock ArXiv e-prints  (2015), 1510.00635.

\bibitem{magic}
J. {Aleksi{\'c}} et~al.,
\newblock Astroparticle Physics 72 (2016) 76, 1409.5594.

\bibitem{hessI}
J.A. {Hinton} and . {the HESS Collaboration},
\newblock New Astronomy 48 (2004) 331, astro-ph/0403052.

\bibitem{veritas}
J. {Holder} et~al.,
\newblock Astroparticle Physics 25 (2006) 391, astro-ph/0604119.

\bibitem{veritas135}
J. {Holder} and . {for the VERITAS Collaboration},
\newblock ArXiv e-prints  (2016), 1609.02881.

\bibitem{hessPC}
C. van Eldik~for HESS~collaboration,
\newblock personal communication, 2016.

\bibitem{hess50gev}
C. {Stegmann} and . {H.~E.~S.~S.~Collaboration},
\newblock American Institute of Physics Conference Series, edited by F.A.
  {Aharonian}, W. {Hofmann} and F.M. {Rieger}, , American Institute of Physics
  Conference Series Vol. 1505, pp. 194--201, 2012.

\bibitem{expacta2011}
M. {Actis} et~al.,
\newblock Experimental Astronomy 32 (2011) 193, 1008.3703.

\bibitem{konopelko}
F.A. {Aharonian} et~al.,
\newblock Astroparticle Physics 15 (2001) 335, astro-ph/0006163.

\bibitem{chinmayB}
C. Borwankar et~al.,
\newblock Astroparticle Physics 84 (2016) 97 .

\bibitem{crabHEGRA}
F.A. {Aharonian} et~al.,
\newblock ApJ 539 (2000) 317, astro-ph/0003182.

\bibitem{stalin}
C.S. {Stalin} et~al.,
\newblock Bulletin of the Astronomical Society of India 36 (2008) 111,
  0809.1745.

\bibitem{Hillas}
A.M. {hillas},
\newblock International Cosmic Ray Conference, edited by . {F.~C.~Jones}, ,
  International Cosmic Ray Conference Vol.~3, pp. 445--448, 1985.

\bibitem{cleaning}
. {HEGRA Collaboration} et~al.,
\newblock Astroparticle Physics 4 (1996) 199.

\bibitem{mradul1}
M. {Sharma} et~al.,
\newblock Research in Astronomy and Astrophysics 14 (2014) 1491, 1410.5125.

\bibitem{magicRF}
J. {Albert} and . {and co-authors}.,
\newblock Nuclear Instruments and Methods in Physics Research A 588 (2008) 424,
  0709.3719.

\bibitem{spectraCosmic1}
S. {Haino} et~al.,
\newblock Physics Letters B 594 (2004) 35, astro-ph/0403704.

\bibitem{spectraCosmic2}
B. {Wiebel-Sooth}, P.L. {Biermann} and H. {Meyer},
\newblock aap 330 (1998) 389, astro-ph/9709253.

\bibitem{magic1}
J. {Aleksi{\'c}} et~al.,
\newblock Astroparticle Physics 35 (2012) 435, 1108.1477.

\bibitem{2fhl}
M. {Ackermann} et~al.,
\newblock ApJS 222 (2016) 5.

\bibitem{time}
E. {Aliu} et~al.,
\newblock Astroparticle Physics 30 (2009) 293, 0810.3568.

\end{thebibliography}

\section{Appendix}
\begin{figure}[!h]
\begin{center}
\includegraphics[height=0.30\textwidth, angle=270]{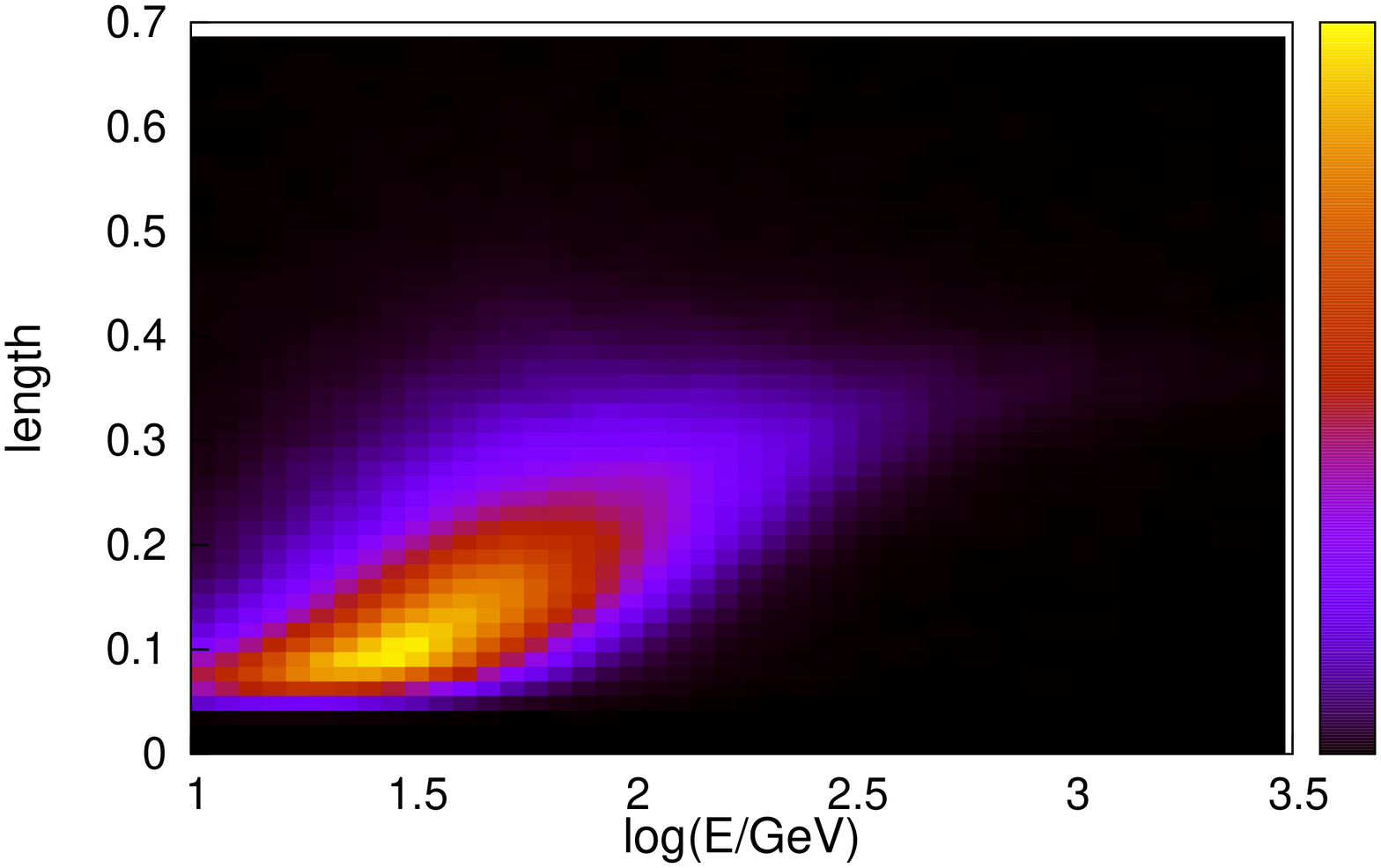}\includegraphics[height=0.30\textwidth, angle=270]{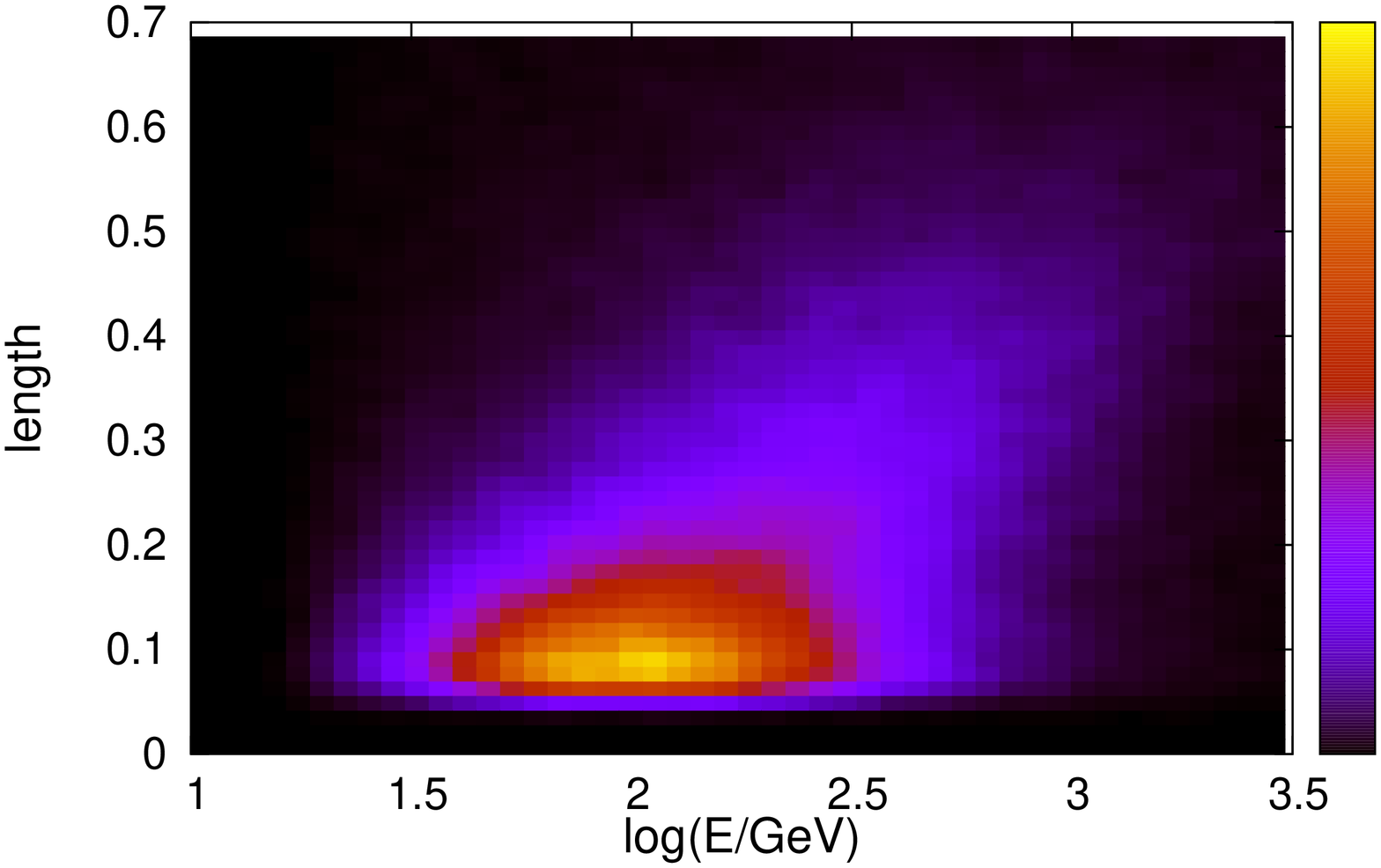}
\includegraphics[height=0.30\textwidth, angle=270]{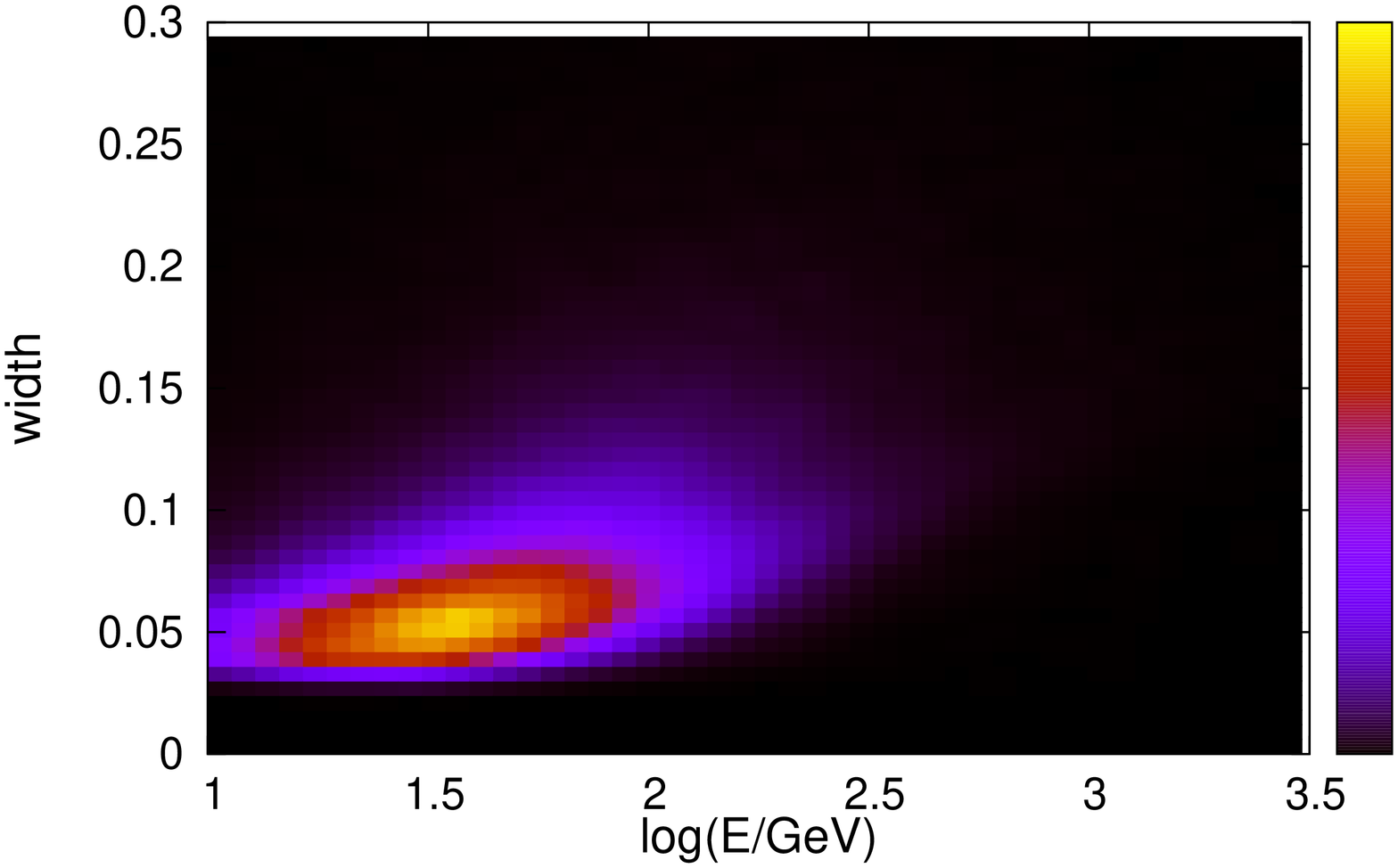}\includegraphics[height=0.30\textwidth, angle=270]{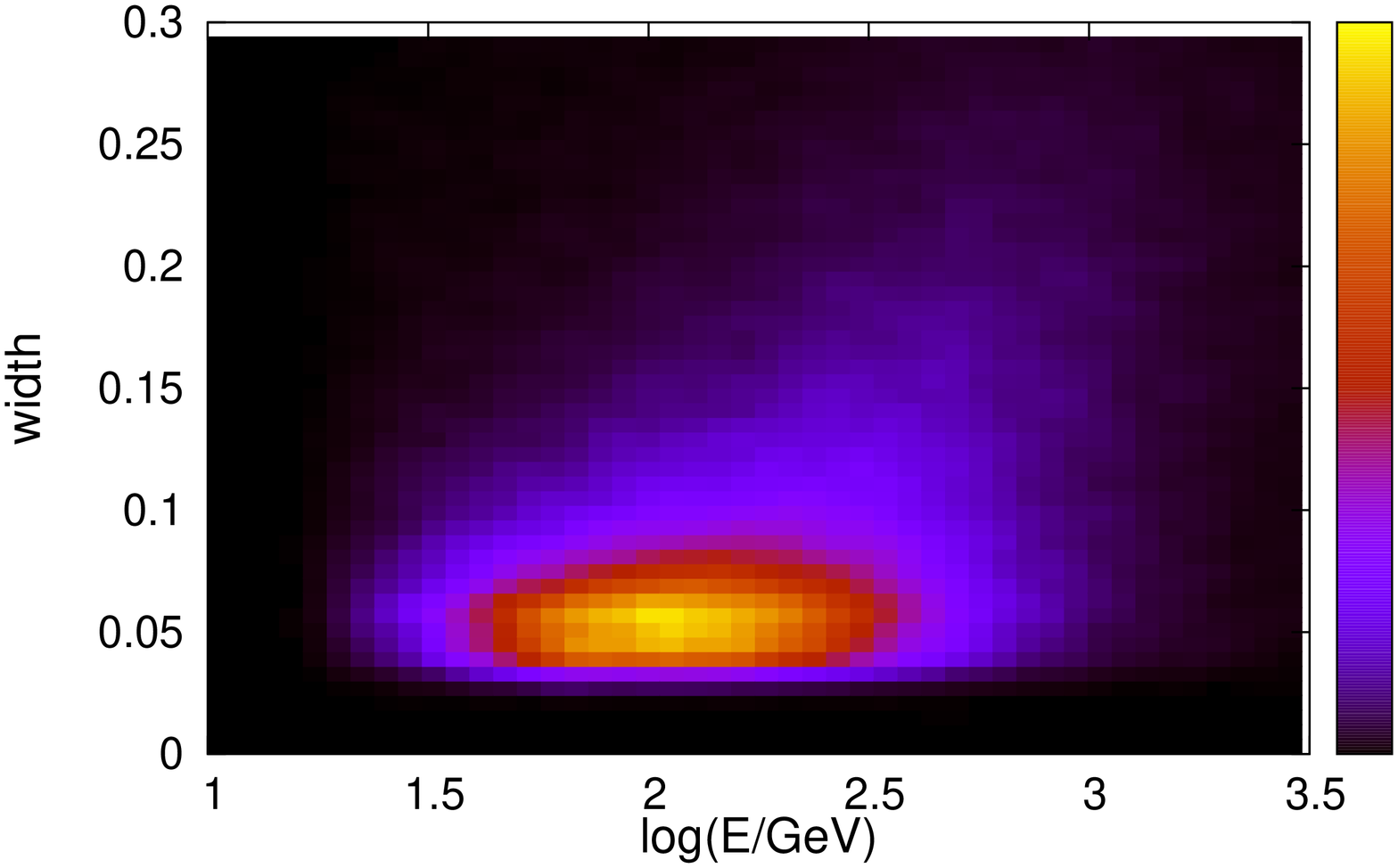}
\caption{\label{Figure.}The length and width distribution of $\gamma$ and proton with Energy. The top panel shows the length vs energy distribution of $\gamma$ and protons whereas the lower panel shows it 
for width vs Energy.}\label{Figure:lwedist}
\end{center}
\end{figure}

The Length and width distribution of $\gamma$ and protons with Energy is shown in Figure \ref{Figure:lwedist}

Sources selected from the second Fermi high energy catalogue visible to the MACE telescope are listed below.
\begin{table}
	\centering
	\begin{tabular}{|l|c|c|c|c|}
		\hline
		Source Name & Energy flux & Spectral & Associated & Source  \\
		            & (ergs/cm$^2$/s) & index  & source     & Class \\
		\hline
2FHLJ1713.5-3945e & 4.840E-011 & 2.03 & RXJ1713.7-3946 & snr \\
2FHLJ1745.1-3035 & 2.689E-011 & 1.25 & -- & unk \\
2FHLJ1745.7-2900 & 2.449E-011 & 2.33 & GalCentreRidge & spp \\
2FHLJ1801.3-2326e & 2.529E-011 & 2.55 & W28 & snr \\
2FHLJ1805.6-2136e & 8.299E-011 & 1.98 & W30 & spp \\
2FHLJ1917.7-1921 & 1.719E-011 & 2.44 & 1H1914-194 & bll \\
2FHLJ1824.5-1350e & 2.400E-010 & 1.89 & HESSJ1825-137 & pwn \\
2FHLJ0349.3-1158 & 1.4700E-011 & 0.95 & 1ES0347-121 & bll \\
2FHLJ1834.5-0846e & 3.810E-011 & 2.27 & W41 & spp \\
2FHLJ1834.6-0701 & 1.779E-011 & 2.25 & -- & unk \\
2FHLJ1836.5-0655e & 5.560E-011 & 2.03 & HESSJ1837-069 & pwn \\
2FHLJ1840.9-0532e & 1.219E-010 & 2.00 & HESSJ1841-055 & pwn \\
2FHLJ0811.9+0238 & 7.039E-012 & 1.61 & PMNJ0811+0237 & bll \\
2FHLJ1923.2+1408e & 1.120E-011 & 3.76 & W51C & snr \\
2FHLJ0648.6+1516 & 1.699E-011 & 2.00 & RXJ0648.7+1516 & bll \\
2FHLJ0319.7+1849 & 1.210E-011 & 1.45 & RBS0413 & bll-g \\
2FHLJ0534.5+2201 & 3.520E-010 & 2.13 & Crab & pwn \\
2FHLJ0617.2+2234e & 4.970E-011 & 2.66 & IC443 & snr \\
2FHLJ0809.5+3458 & 1.089E-011 & 1.09 & B20806+35 & bll-g \\
2FHLJ2016.2+3713 & 9.530E-012 & 1.74 & SNRG74.9+1.2 & spp \\
2FHLJ1104.4+3812 & 3.290E-010 & 2.14 & Mkn421 & bll \\
2FHLJ2249.9+3826 & 1.410E-011 & 1.68 & B32247+381 & bll \\
2FHLJ1653.9+3945 & 1.280E-010 & 2.13 & Mkn501 & bll \\
2FHLJ2021.0+4031e & 6.719E-011 & 1.99 & GammaCygni & snr \\
2FHLJ0316.6+4120 & 1.330E-011 & 1.34 & IC310 & rdg \\
2FHLJ1015.0+4926 & 3.300E-011 & 2.50 & 1H1013+498 & bll \\
2FHLJ2056.7+4939 & 1.100E-011 & 2.33 & RGBJ2056+496 & bcuII \\
2FHLJ2347.1+5142 & 2.620E-011 & 1.85 & 1ES2344+514 & bll \\
2FHLJ0048.0+5449 & 7.640E-012 & 1.30 & 1RXSJ004754.5+544758 & bcuII \\
2FHLJ0431.2+5553e & 4.939E-011 & 1.66 & SNRG150.3+4.5 & snr \\
2FHLJ2323.4+5848 & 1.640E-011 & 2.45 & CasA & snr \\
2FHLJ2000.1+6508 & 5.259E-011 & 1.89 & 1ES1959+650 & bll \\
2FHLJ0507.9+6737 & 4.319E-011 & 2.15 & 1ES0502+675 & bll \\
2FHLJ0153.5+7113 & 5.059E-012 & 1.61 & TXS0149+710 & bcuI \\
		\hline
	\end{tabular}
	\caption{Sources, selected from second Fermi high energy catalogue, which are visible
	at MACE telescope site and satisfies the criteria defined in the text}
	\label{Table:2fhl}
\end{table}

\end{document}